\documentclass[12pt,english,english,english,english,english,english]{article}
\usepackage[T1]{fontenc}
\usepackage[latin1]{inputenc}
\usepackage{graphicx}
\usepackage{amssymb}

\makeatletter

\providecommand{\LyX}{L\kern-.1667em\lower.25em\hbox{Y}\kern-.125emX\@}

\usepackage[T1]{fontenc}
\usepackage[latin1]{inputenc}
\usepackage{babel}

\makeatletter

\usepackage[T1]{fontenc}
\usepackage[latin1]{inputenc}
\usepackage{babel}

\makeatletter

\usepackage[T1]{fontenc}
\usepackage[latin1]{inputenc}
\usepackage{babel}

\makeatletter

\usepackage[T1]{fontenc}
\usepackage[latin1]{inputenc}
\usepackage{babel}

\makeatletter

\usepackage[T1]{fontenc}
\usepackage[latin1]{inputenc}
\usepackage{babel}



\makeatother

\makeatother

\makeatother

\makeatother

\usepackage{babel}
\makeatother
\begin{document}

\title{Entropy bounds for massive scalar field in positive curvature space}

\author{{ E. Elizalde}%
\footnote{e-mail: elizalde@ieec.fcr.es%
}\\
\small Consejo Superior de Investigaciones Cient\'{\i}ficas \\
\small Institut d'Estudis Espacials de Catalunya (IEEC/CSIC) \\
\small  Edifici Nexus, Gran Capit\`{a} 2-4, 08034 Barcelona; \ and \\
\small  Departament ECM,  Facultat de F\'{\i}sica, Universitat de Barcelona \\
\small  Av. Diagonal 647, 08028 Barcelona, Spain \\
\\
{A. C. Tort}%
\footnote{e-mail: tort@if.ufrj.br. Present address: Institut d'Estudis Espacials
de Catalunya (IEEC/CSIC) \hspace*{5mm} Edifici Nexus, 
Gran Capit\`{a} 2-4, 08034
Barcelona, Spain; e-mail address: visit11@ieec.fcr.es%
}\\
\small  Departamento de F\'{\i}sica Te\'{o}rica - Instituto de F\'{\i}sica
\\
\small Universidade Federal do Rio de Janeiro\\
\small  Caixa Postal 68.528; CEP 21941-972 Rio de Janeiro, Brazil}

\maketitle

\begin{abstract}
\noindent We consider a massive scalar field with arbitrary coupling
in $\mathbf{S}^{1}\times \mathbf{S}^{3}$ space, which mimics the
thermal expanding universe, and calculate explicitly all relevant 
thermodynamical
functions in the low- and high-temperature regimes, extending previous
analysis of entropy bounds and entropy/energy ratios performed in the conformal
case. For high temperatures, new mass-dependent
entropy ratios are established which, differently to the conformal limit,
 fulfil Bekenstein's and  Verlinde's bounds in the physical region. \\ 
\\
PACS numbers 04.62,+v; 11.10.Wx; 98.80.Hw
\end{abstract}

\section{Introduction}

According to recent astrophysical data our universe goes through (or
will go through, sooner or later) a de Sitter phase. Moreover, 
according to the inflationary
paradigm%
\footnote{Notice however that there is also the point of view due to 
Ellis and Maartens
\cite{Ellis&Maarten02} which contemplates the possibility of an eternal
inflationary universe without quantum gravity.%
}, the universe in the past was also a de Sitter space. Taking into
consideration the fact that the early universe was thermal, one of
the most promising candidates for modelling the early universe is
the $\mathbf{S}^{1}\times \mathbf{S}^{3}$ space.

One of the fundamental issues about the early universe is the entropy
issue, particularly, the occurence of the so called entropy bounds.
One of the better known of these bounds involves the ratio of the entropy
$S$ to the total energy $E$ of a closed physical system at high
temperature, the Bekenstein bound \cite{Bekenstein81}. For $\mathbf{S}^{1}\times \mathbf{S}^{3}$
space the Bekenstein bound reads \cite{Brustein01}\begin{equation}
\frac{S}{2\pi rE}\leq 1,\end{equation}
where $r$ is the radius associated with $\mathbf{S}^{3}$. Recently,
however, a more restringent form of the above bound was proposed by
Verlinde \cite{Verlinde} (for further developments of the Verlinde
formalism in four-dimensional de Sitter space see \cite{Abdalla&01}).
The Verlinde bound, which applies only to conformal field theories
(CFTs) and admits an holographic interpretation, is given by\begin{equation}
\frac{S}{2\pi rE}\leq \frac{1}{3}.\end{equation}
The explicit verification of the Verlinde bound was performed for
a number of CFTs, mainly in $\mathbf{S}^{1}\times \mathbf{S}^{3}$
space, via the calculation of the corresponding partition function
\cite{Abdalla&01,Kutasov and Larsen 2000,Klemm at al 2000,Lin2000,Nojori2000}.
The corresponding quantum bounds are weaker than the Verlinde CFT
bound. As it should be expected, entropy bounds occur only for high
temperature; in the low temperature regime the corresponding entropy/energy
ratio can be very large, in fact, it is limitless \cite{Brevik al 02}.
For the early universe, however, only the high temperature regime
matters. This being so, the fundamental question is: What is the most
realistic entropy bound for the early universe? 

So far, calculations of quantum versions of entropy bounds have been
limited to CFT. For example, the cases of a conformally invariant
massless scalar field in spacetime of dimension $D=4$, and of other
types of massless fields in $\mathbf{S}^{1}\times \mathbf{S}^{3}$
space, were considered in \cite{Brevik al 02} where the Casimir energies
\cite{Milton2001} at zero and finite temperature were exactly calculated.
The purpose of the present paper is to address this question, in the
framework of non-conformal field theory, by considering a simple
example: the case of a massive scalar field with an arbitrary
scalar-gravitational coupling in $\mathbf{S}^{1}\times \mathbf{S}^{3}$
space. It seems clear that this is a necessary step in the way towards
 the calculation
of entropy bounds for  more realistic quantum gravity theories (for a 
general introduction see \cite{BOS}) in the early
de Sitter universe, where gravitons are massive.

The paper is divided as follows. In Sect. 2 we briefly sketch a general
zeta function formalism suitable for $\mathbf{S}^{1}\times \mathbf{S}^{d}$
geometries, which is fundamented in previous work. In Sect. 3, we apply 
this formalism to the special case
of the $\mathbf{S}^{1}\times \mathbf{S}^{3}$ geometry and obtain
the low and high temperature representations for the basic set of
 thermodynamical
functions. In Sect. 4, the implication of the results for what concerns
the entropy bounds is discussed. The last section contains
a concluding summary. 

Throughout the paper, purely thermal quantities are denoted with a
tilde, thus $\widetilde{E}$ stands for the thermal energy, while
untilded quantities include zero point contributions. We also employ
natural units, $\hbar =c=1$, and set the Boltzmann constant equal
to one.

\section{Zeta function formalism}

We begin by reviewing the generalised zeta function approach \cite{Elisalde94}
for the construction of the logarithm of the partition function of
the problem at hand. The eigenvalue equation in $\mathbf{S}^{1}\times \mathbf{S}^{d}$
reads\begin{equation}
\left(-\partial _{d}^{2}+\xi R+m^{2}\right)\Phi _{\lambda }\left(\tau ,\vec{x}\right)=\lambda \Phi _{\lambda }\left(\tau ,\vec{x}\right),
\label{main eigenvalue eq}\end{equation}
with $\partial _{d}^{2}\equiv \partial ^{2}/\partial \tau ^{2}+\vec{\nabla }_{d}^{2}$,
where $\tau :=it$ is the euclidean time coordinate and $\vec{\nabla }_{d}^{2}$
 the Laplace operator on $\mathbf{S}^{d}$. The parameter $\xi $
is the conformal parameter, whose  value is a function of
the spacetime dimensions $D=d+1$: $\left(D-2\right)/\left[4\left(D-1\right)\right]$;
 $R$ is the Ricci curvature scalar. Equation (\ref{main eigenvalue eq})
splits into two eigenvalue equations. The first is
\begin{equation}
\left(-\frac{\partial ^{2}}{\partial \tau ^{2}}+m^{2}\right)\Omega \left(\tau \right)=\alpha \Omega \left(\tau \right),\label{tau eigenvalue eq}\end{equation}
 where $\Omega \left(\tau \right)$ satisfies periodic boundary conditions
on $\mathbf{S}^{1}$, so that $\Omega \left(\tau \right)=\Omega \left(\tau +\beta \right)$,
where $\beta \equiv T^{-1}$, the reciprocal of the temperature $T$, is
the periodicity length for the compactified euclidean time axis. Oscillating
solutions of Eq. (\ref{tau eigenvalue eq}) are given by $\Omega \left(\tau \right)\sim \exp \left(ik_{\tau }\tau \right)$
with $k_{\tau }=2n\pi /\beta $, where $n$ is an integer. It then follows,
 that \begin{equation}
\alpha \rightarrow \alpha _{n}=\left(\frac{2\pi n}{\beta }\right)^{2}+m^{2},\end{equation}
and \begin{equation}
\Omega \left(\tau \right)\rightarrow \Omega _{n}\left(\tau \right)\sim \exp \left(i2n\pi /\beta \right),\, \, \, \, n=0,\pm 1,\pm 2,\ldots .\end{equation}
The second eigenvalue equation, steming from Eq. (\ref{main eigenvalue eq}),
 for the Laplacian on $S^{d}$ is \begin{equation}
\left(-\vec{\nabla }_{d}^{2}+\xi R\right)\Psi _{\nu }\left(\vec{x}\right)=\nu \Psi _{\nu }\left(\vec{x}\right).\end{equation}
The eigenfunctions $\Psi _{\nu }\left(\vec{x}\right)$ can be chosen
as the spherical harmonics $Y_{\ell m}\left(\theta _{1},\theta _{2},...\theta _{d-1},\phi \right)$
on $S^{d}$, which satisfy  \begin{equation}
-\vec{\nabla }_{d}^{2}Y_{\ell m}\left(\vec{x}\right)=\frac{M_{\ell }^{2}}{r^{2}}Y_{\ell m}\left(\vec{x}\right),\end{equation}
and it follows that\begin{equation}
\nu \rightarrow \nu _{\ell }=\frac{M_{\ell }^{2}}{r^{2}}+\xi R.\end{equation}
 The complete spectrum of eigenvalues defined by the eigenvalue problem in
$\mathbf{S}\times \mathbf{S}^{d}$, expressed by Eq. (\ref{main eigenvalue eq}),
is then\begin{equation}
\lambda \rightarrow \lambda _{n\ell }=\alpha _{n}+\nu _{\ell }=\left(\frac{2\pi n}{\beta }\right)^{2}+m^{2}+\xi R+\frac{M_{\ell }^{2}}{r^{2}}.\end{equation}
The zeta function corresponding to the operator $-\partial _{d}^{2}+\xi R+m^{2}$
is given by\begin{equation}
\zeta \left(s \left| \frac{-\partial _{d}^{2}+\xi R+m^{2}}{\sigma ^{2}}\right.
\right)=\sigma ^{2s}\sum _{n=-\infty }^{\infty }\sum _{\ell =0}^{\infty }\left[\left(\frac{2\pi n}{\beta }\right)^{2}+\frac{M_{\ell }^{2}\left(\mu ,\chi \right)}{r^{2}}\right]^{-s},\end{equation}
where $\sigma$ is a regularization parameter with dimensions of mass 
\cite{Elisalde94}
 and $M_{\ell }^{2}\left(\mu ,\chi \right)$ is defined as \begin{equation}
M_{\ell }^{2}\left(\mu ,\chi \right)\equiv M_{\ell }^{2}+\mu ^{2}+\chi ,
\end{equation}
with $\mu \equiv mr$ and $\chi =\xi Rr^{2}$. For the bosonic scalar
field we are dealing with, the partition function can be obtained
from \begin{equation}
\log \, Z\left(\beta \right)=\frac{1}{2}\frac{d}{ds}\zeta \left(s=0\left|
\frac{-\partial _{d}^{2}+\xi R+m^{2}}{\sigma ^{2}}\right.\right).
\label{partion function I}\end{equation}
Now, we proceed along the lines described in Ref. \cite{Elizalde&Tort2002},
i.e., we introduce the Mellin transform and make use of the Poisson
sum formula (or the Jacobi theta function identity) to formally split
the partition function into two separate contributions, one corresponding
to the zero-temperature Casimir energy and one corresponding to its
finite temperature correction. The final result is

\begin{equation}
\log \, Z\left(\beta \right)=-\frac{\beta }{2r}\sum _{\ell =0}^{\infty }D_{\ell }M_{\ell }\left(\mu ,\chi \right)-\sum _{\ell =0}^{\infty }D_{\ell }\log \left(1-e^{-\beta M_{\ell }\left(\mu ,\chi \right)/r}\right),\label{partition function II}\end{equation}
Equation (\ref{partition function II}) holds generally for $\mathbf{S}^{1}\times \mathbf{S}^{d}$
geometries and contains all the relevant physical information we need
at zero and finite temperature. The case of spatial dimension $d=1$,
however, needs to be handled with special care: the formalism must
be slightly modified in order to take into account the presence of a zero mode.
In fact, it can be shown that the thermodynamics of the zero mode
are internally inconsistent with the choice of a particular value for
the scaling mass, they thus turn out to be unpredictive and, therefore, must
be discarded from the formalism (see Ref. \cite{Elizalde&Tort2002}
for more details and Ref. \cite{Brevik al 02} for an independent
statistical mechanical discussion). The simplicity, together with the
suggestive physical meaning
of Eq. (\ref{partition function II}), which provides a neat formal separation
of the logarithm of the partition function into non-thermal and thermal
sectors, is a direct, remarkable consequence of the generalised zeta-function
technique employed here.

\section{The massive scalar field on $\mathbf{S}^{1}\times \mathbf{S}^{3}$:
Thermodynamical functions for the low- and high-T regimes}

We now apply the formalism described above to our specific problem,
i.e.: the evaluation of the Casimir zero point energy and the thermodynamics
of a massive scalar field $\Phi $ in a $\mathbf{S}^{1}\times \mathbf{S}^{3}$
geometry, with an explicit conformal symmetry breaking due to mass,
 and deviations
from the conformal value $\xi =1/6$. The eigenvalues of the Laplace
operator on $\mathbf{S}^{3}$ are given by $M_{\ell }^{2}=\ell \left(\ell +1\right)$,
with $\ell =0,1,2,\ldots$, and degeneracy $D_{\ell }=\left(\ell +1\right)^{2}$.
Hence, we can write \begin{equation}
M_{\ell }^{2}\left(\mu ,\chi \right)\equiv \left(\ell +1\right)^{2}+\mu _{eff}^{2}\end{equation}
where the dimensionless parameter $\mu _{eff}^{2}$ is defined by
\begin{equation}
\mu _{eff}^{2}\equiv \mu _{eff}^{2}\left(\mu ,\chi \right)=\mu ^{2}+\chi -1.\label{effective mass}\end{equation}
This parameter plays a role similar to an effective mass. Remark,
however, that $\mu _{eff}^{2}$ and $\chi $ are real numbers
and that $\mu ^{2}\geq 0$. The standard conformal value ($\xi =1/6$) 
corresponds
to $\mu ^{2}=0$ and $\chi =1$ ($\mu _{eff} =0$). 

The Casimir energy at zero temperature, which can be formally read
out from Eq. (\ref{partition function II}), is given by\begin{equation}
E_{0}\left(\mu _{eff}^{2}\right)=\frac{1}{2r}\sum _{\ell =0}^{\infty }\left(\ell +1\right)^{2}\sqrt{\left(\ell +1\right)^{2}+\mu _{eff}^{2}}.\label{zero-point energy}\end{equation}
 In the conformal limit, we obtain\begin{equation}
E_{0}\left(\mu _{eff}^{2}\left(\mu ^{2}=0,\chi =1\right)\right)=\frac{1}{2r}\zeta \left(-3\right)=\frac{1}{240r}.\end{equation}
 The Casimir energy for arbitrary $\mu _{eff}^{2}$ can be calculated
following the analytical methods discussed in \cite{Elisalde94}.
In the following subsections we will consider,  however, an alternative
way.

\subsection{Thermodynamical functions for low T}

Let us now focus on the thermal sector, which is described by the
second term on the r.h.s. of Eq. (\ref{partition function II}), and
obtain the relevant thermodynamical functions. For convenience, we
discuss the low and the high temperature regimes separately, and for
the latter case, we distinguish between the open intervals $0<\mu _{eff}^{2}<\infty $
and $-1<\mu _{eff}^{2}<0$. After a convenient shift of the summation
index, Eq. (\ref{partition function II}) reads \begin{equation}
\log \widetilde{Z}\left(\beta \right)=-\sum _{\ell =1}^{\infty }\ell ^{2}\log \left(1-e^{-\beta \sqrt{\ell ^{2}+\mu _{eff}^{2}}/r}\right).\label{thermal pf}\end{equation}
The thermal energy is\begin{equation}
\widetilde{E}\left(\beta \right)=-\frac{d}{d\beta }\log \widetilde{Z}\left(\beta \right)=\frac{1}{r}\sum _{\ell =1}^{\infty }\frac{\ell ^{2}\left(\ell ^{2}+\mu _{eff}^{2}\right)^{1/2}}{e^{\beta \left(\ell ^{2}+\mu _{eff}^{2}\right)^{1/2}/r}-1},\label{Thermal energy}\end{equation}
and the thermal part of the free energy 
\begin{equation}
\widetilde{F}\left(\beta \right)=-\frac{1}{\beta }\log \widetilde{Z}\left(\beta \right)=-\frac{1}{\beta }\sum _{\ell =1}^{\infty }\ell ^{2}\log \left(1-e^{-\beta \sqrt{\ell ^{2}+\mu _{eff}^{2}}/r}\right). \end{equation}
 The full set of thermodynamical functions read \begin{equation}
E\left(\beta \right)=E_{0}+\widetilde{E}\left(\beta \right)=\frac{1}{2r}\sum _{\ell =1}^{\infty }\ell ^{2}\left(\ell ^{2}+\mu _{eff}^{2}\right)^{1/2}+\frac{1}{r}\sum _{\ell =1}^{\infty }\frac{\ell ^{2}\left(\ell ^{2}+\mu _{eff}^{2}\right)^{1/2}}{e^{\beta \left(\ell ^{2}+\mu _{eff}^{2}\right)^{1/2}/r}-1},\end{equation}
\begin{equation}
F\left(\beta \right)=F_{0}+\widetilde{F}\left(\beta \right)=\frac{1}{2r}\sum _{\ell =1}^{\infty }\ell ^{2}\left(\ell ^{2}+\mu _{eff}^{2}\right)^{1/2}+\frac{1}{\beta }\sum _{\ell =1}^{\infty }\ell ^{2}\log \left[1-e^{-\beta \left(\ell ^{2}+\mu _{eff}^{2}\right)^{1/2}/r}\right]\end{equation}
\begin{equation}
S\left(\beta \right)=\beta \left(E-F\right)=\frac{\beta }{r}\sum _{\ell =1}^{\infty }\frac{\ell ^{2}\left(\ell ^{2}+\mu _{eff}^{2}\right)^{1/2}}{e^{\beta \left(\ell ^{2}+\mu _{eff}^{2}\right)^{1/2}/r}-1}-\sum _{\ell =1}^{\infty }\ell ^{2}\log \left[1-e^{-\beta \left(\ell ^{2}+\mu _{eff}^{2}\right)^{1/2}/r}\right].\end{equation}
Notice that when $\beta \rightarrow \infty $ the entropy goes to
zero as it should. It is clear that the zeta function approach naturally
leads to low-temperature representations for the fundamental thermodynamical
quantities.

\subsubsection{The low-T approximation}

In the low-T limit, $\beta /r\gg 1$, we can easily obtain, from the
low-T representations above, the following expressions for $\widetilde{E}\left(\beta \right)$,
$\widetilde{F}\left(\beta \right)$, and $S\left(\beta \right)$\begin{equation}
\widetilde{E}\left(\beta \right)\approx \frac{\left(1+\mu _{eff}^{2}\right)^{1/2}}{r}e^{-\beta \left(1+\mu _{eff}^{2}\right)^{1/2}/r},\label{lowTenergy}\end{equation}
\begin{equation}
\widetilde{F}\left(\beta \right)\approx -\frac{1}{\beta }e^{-\beta \left(1+\mu _{eff}^{2}\right)^{1/2}/r},\label{lowTfreeenergy}\end{equation}
\begin{equation}
S\left(\beta \right)\approx \frac{\beta \left(1+\mu _{eff}^{2}\right)^{1/2}}{r}e^{-\beta \left(1+\mu _{eff}^{2}\right)^{1/2}/r}.\label{lowTentropy}\end{equation}
As we will see, these relations lead to simple entropy bounds, similar
to those obtained in the conformal case.

\subsection{Thermodynamical functions for the high-T regime }

In order to construct high-temperature representations for the relevant
thermodynamical functions, we can make use of one of the several well
known summation formulae. The most convenient one here is the rescaled
Abel-Plana formula \cite{Erdelyietal1953}\begin{equation}
h\sum _{\ell =1}^{\infty }f\left(h\ell \right)=-\frac{h}{2}f\left(0\right)+\int _{0}^{\infty }dx\, f\left(x\right)+i\int _{0}^{\infty }\frac{f\left(iy\right)-f\left(-iy\right)}{e^{2\pi y/h}-1},\label{Abel-Plana}\end{equation}
 where $h$ is a scale factor and $f\left(x\right)$ is such that
$f\left(\infty \right)=0$. In the high temperature regime we must
distinguish between two cases, to wit: $0<\mu _{eff}^{2}<\infty $
and $-1<\mu _{eff}^{2}<0$. Let us begin with the first.

\subsubsection{Thermodynamical functions for $0<\mu _{eff}^{2}<$$\infty $}

Consider the open interval $0<\mu _{eff}^{2}<\infty $ and the purely
thermal part of the logarithm of the partition function, Eq. 
(\ref{thermal pf}).
The rescaled Abel-Plana formula with the identification $h=\beta /r$
and $f\left(x\right)=-\left(r/\beta \right)^{3}x^{2}\log 
\left[1-e^{-\left(x^{2}+m_{eff}^{2}\beta ^{2}\right)^{1/2}}\right]$,
where $m _{eff} =\mu _{eff}/r$ is the effective mass corresponding to the 
dimensionless parameter $\mu _{eff}$,
reads\begin{eqnarray}
\frac{\beta }{r}\sum _{\ell =1}^{\infty }f\left(\frac{\beta }{r}\ell \right)
&=&-\left(\frac{r}{\beta }\right)^{3}\int _{0}^{\infty }dx\, x^{2}\log \left[1-e^{-\left(x^{2}+m_{eff}^{2}\beta ^{2}\right)^{1/2}}\right]  \nonumber \\
&&-2\left(\frac{r}{\beta }\right)^{3}\, \sum _{k=1}^{\infty }\frac{1}{k}\int _{m_{eff}}^{\infty }dy\, y^{2}\frac{\sin \left[k\left(y^{2}-m_{eff}^{2}\beta ^{2}\right)^{1/2}\right]}{e^{2\pi ry/\beta }-1}.  \label{Abel-Plana 2}
\end{eqnarray}
Let us consider the first term on the r.h.s. of Eq. (\ref{Abel-Plana 2}).
Introducing a new real, positive variable $t$, defined by $t\equiv +\left(x^{2}+m_{eff}^{2}\beta ^{2}\right)^{1/2}$,
and expanding the logarithm, we readily obtain\begin{eqnarray}
\int _{0}^{\infty }dx\, x^{2}\log \left[1-e^{-\left(x^{2}+m_{eff}^{2}\beta ^{2}\right)^{1/2}}\right]=-\sum _{n=1}^{\infty }\frac{1}{n}\int _{m_{eff}}^{\infty }dt\, t\, \left(t^{2}-m_{eff}^{2}\beta ^{2}\right)^{1/2}e^{-nt} &  & \nonumber \\
=\frac{d}{d\lambda }\sum _{n=1}^{\infty }\frac{1}{n^{2}}\int _{m_{eff}}^{\infty }dt\, \left(t^{2}-m_{eff}^{2}\beta ^{2}\right)^{1/2}e^{-nt\lambda }\vert _{\lambda =1}. &  & 
\end{eqnarray}
The integral can be obtained with the help of (\emph{cf.} Ref. \cite{Grad94},
formula 3.887.6)\begin{equation}
\int _{\mu _{1}}^{\infty }dx\, \left(x^{2}-\mu _{1}^{2}\right)^{\nu -1}e^{-\mu _{2}x}=\frac{1}{\sqrt{\pi }}\left(\frac{2\mu _{1}}{\mu _{2}}\right)^{\nu -1/2}\Gamma \left(\nu \right)K_{\nu -\frac{1}{2}}\left(\mu _{1}\mu _{2}\right),\label{grad int}\end{equation}
where $K_{\nu }\left(z\right)$ is the modified Bessel function of
the third kind. This result holds for $\mu _{1}>0;\, \, \Re \, \mu _{2}>0;\, \, \Re \, \nu >0$.
It follows that\begin{equation}
\int _{0}^{\infty }dx\, x^{2}\log \left[1-e^{-\left(x^{2}+m_{eff}^{2}\beta ^{2}\right)^{1/2}}\right]=m_{eff}\beta \sum _{n=1}^{\infty }\frac{1}{n^{3}}\, \left[\frac{d}{d\lambda }\, \frac{1}{\lambda }\, K_{1}\left(m_{eff}\beta \lambda n\right)\right]_{\lambda =1}.\end{equation}
Now, we make use of the relation (\emph{cf}. Ref. \cite{Grad94}, formula
8.486.15)\begin{equation}
\left(\frac{d}{z\, dz}\right)^{m}\left[z^{-\nu }K_{\nu }\left(z\right)\right]=\left(-1\right)^{m}z^{-\nu -m}K_{\nu +m}\left(z\right),\label{recursion 1}\end{equation}
to obtain, after simple manipulations, \begin{equation}
\int _{0}^{\infty }dx\, x^{2}\log \left[1-e^{-\left(x^{2}+m_{eff}^{2}\beta ^{2}\right)^{1/2}}\right]=-\left(m_{eff}\beta \right)^{2}\, \sum _{n=1}^{\infty }\frac{1}{n^{2}}K_{2}\left(m_{eff}\beta n\right).\end{equation}
Let us now consider the second term in Eq. (\ref{Abel-Plana 2}).
Expanding the sine function in its integrand,
according to \begin{equation}
\sin \left[k\left(y^{2}-m_{eff}^{2}\beta ^{2}\right)^{1/2}\right]=\sum _{n=1}^{\infty }\frac{\left(-1\right)^{n-1}}{\left(2n-1\right)!}\left(y^{2}-m_{eff}^{2}\beta ^{2}\right)^{n-1/2}k^{2n-1},\label{sinetaylor}\end{equation}
we obtain an infinite sum of convergent integrals which, after performing
the sum over $k$, yields \begin{equation}
i\int _{0}^{\infty }\frac{f\left(iy\right)-f\left(-iy\right)}{e^{2\pi y/h}-1}=-2\left(\frac{r}{\beta }\right)^{3}\, \sum _{n=1}^{\infty }\frac{\left(-1\right)^{n-1}}{\left(2n-1\right)!}\, \zeta \left(2-2n\right)G_{n}\left(m_{eff}\beta \right),\label{AB-thired term 2}\end{equation}
where we have defined\begin{equation}
G_{n}\left(m_{eff}\beta \right)\equiv \int _{m_{eff}\beta }^{\infty }dy\, y^{2}\frac{\left(y^{2}-m_{eff}^{2}\beta ^{2}\right)^{n-1/2}}{e^{2\pi ry/\beta }-1}.\end{equation}
Since the Riemann zeta function is zero for all negative even integers,
we have that $\zeta \left(2-2n\right)=0$ for all $n$, except for $n=1$.
Therefore, only the first term of the sum in Eq. (\ref{AB-thired term 2})
survives and, after collecting all pieces, we obtain for the thermal
part of the logarithm of the partition function, the result\begin{equation}
\log \widetilde{Z}\left(\beta \right)=\left(\frac{r}{\beta }\right)^{3}\left(m_{eff}\beta \right)^{2}\, \sum _{n=1}^{\infty }\frac{1}{n^{2}}K_{2}\left(m_{eff}\beta n\right)-2\left(\frac{r}{\beta }\right)^{3}\zeta \left(0\right)G_{1}\left(m_{eff}\beta \right).\label{thermal partition function}\end{equation}
We can check  Eq. (\ref{thermal partition function}) out by setting
$m_{eff}=0$; then, making use of the small $z$ approximation of
$K_{\nu }\left(z\right)\approx \left(1/2\right)\Gamma \left(\nu \right)\left(z/2\right)^{-\nu }$
and of (\emph{cf}. \cite{Grad94}, formula 3.411.1)\begin{equation}
\int _{0}^{\infty }dx\, \frac{x^{\nu -1}}{e^{ux}-1}=\frac{\Gamma \left(\nu \right)}{u^{\nu }}\zeta \left(\nu \right),\end{equation}
 it follows that\begin{equation}
\log \widetilde{Z}\left(\beta \right)\approx \frac{\pi ^{4}}{45}\left(\frac{r}{\beta }\right)^{3}+\frac{1}{240}\frac{\beta }{r},\end{equation}
which is in perfect 
agreement with Kutasov and Larsen \cite{Kutasov and Larsen 2000}
for the conformal limit. It is important to notice, however, that now
this result can be extended to conformal symmetry breaking fields
$\mathbf{S}^{1}\times \mathbf{S}^{3}$, as long as their {\it effective}
mass remains zero. 

More progress can be made if we evaluate $G_{1}\left(m_{eff}\beta \right)$
for an arbitrary effective mass. This can be done by first expanding
the denominator in $G_{1}\left(m_{eff}\beta \right)$ according to
\begin{equation}
\frac{1}{e^{x}-1}=\sum _{n=1}^{\infty }\, e^{-nx},\label{Denominator}\end{equation}
which then allows us to write\begin{equation}
G_{1}\left(m_{eff}\beta \right)=\left(\frac{\beta }{2\pi }\right)^{2}\frac{d^{2}}{d\lambda ^{2}}\sum _{n=1}^{\infty }\frac{1}{n^{2}}\int _{m_{eff}\beta }^{\infty }dy\left(y^{2}-m_{eff}^{2}\beta ^{2}\right)^{1/2}e^{-2\pi nry\lambda /\beta },\end{equation}
where once more the derivative must be evaluated at $\lambda =1$.
This integral can be calculated exactly, with the help of Eq. (\ref{grad int}),
the result being \begin{equation}
G_{1}\left(m_{eff}\beta \right)=\left(\frac{\beta }{2\pi r}\right)^{3}\, m_{eff}\beta \sum _{n=1}^{\infty }\frac{1}{n^{3}}\frac{d^{2}}{d\lambda ^{2}}\frac{1}{\lambda }K_{1}\left(2\pi n\mu _{eff}\lambda \right).\end{equation}
Now make use of Eq. (\ref{recursion 1}) and of the recursion relation
(see \cite{Grad94}, formula 8.486.13)\begin{equation}
z\frac{dK_{\nu }\left(z\right)}{dz}-\nu K_{\nu }\left(z\right)=-zK_{\nu }\left(z\right),\end{equation}
to obtain the useful formula \begin{equation}
\frac{d^{2}}{d\lambda ^{2}}\left[\frac{1}{\lambda }K_{1}\left(2\pi n\mu _{eff}\lambda \right)\right]_{\lambda =1}=-\left(2\pi n\mu _{eff}\right)K_{2}\left(2\pi n\mu _{eff}\right)+\left(2\pi n\mu _{eff}\right)^{2}K_{3}\left(2\pi n\mu _{eff}\right),\end{equation}
so that the second term in Eq. (\ref{thermal partition function}) reads
\begin{equation}
J_{1}\left(m_{eff},\beta \right)=-\frac{\mu _{eff}^{2}\beta }{\left(2\pi \right)^{2}r}\, \sum _{n=1}^{\infty }\frac{1}{n^{2}}K_{2}\left(2\pi n\mu _{eff}\right)+\frac{\mu _{eff}^{3}\beta }{2\pi r}\sum _{n=1}^{\infty }\frac{1}{n}K_{3}\left(2\pi n\mu _{eff}\right),\end{equation}
where we have  made use of $\zeta \left(0\right)=-1/2$.

 Finally,
collecting all pieces, we obtain\begin{eqnarray}
\log \widetilde{Z}\left(\beta \right)=\frac{r\mu _{eff}^{2}}{\beta }\sum _{n=1}^{\infty }\frac{1}{n^{2}}K_{2}\left(m_{eff}\beta n\right)-\frac{\mu _{eff}^{2}\beta }{\left(2\pi \right)^{2}r}\, \sum _{n=1}^{\infty }\frac{1}{n^{2}}K_{2}\left(2\pi n\mu _{eff}\right) &  & \nonumber \\
+\frac{\mu _{eff}^{3}\beta }{2\pi r}\sum _{n=1}^{\infty }\frac{1}{n}K_{3}\left(2\pi n\mu _{eff}\right), &  & \label{high T partition function I}
\end{eqnarray}
which is the result we were looking for. In order to obtain the full set of 
thermodynamical functions, to this result we must add the zero point
contribution\begin{equation}
\log Z\left(\infty \right)=-\frac{\beta }{2r}\sum _{n=1}^{\infty }n^{2}\left(n^{2}+\mu _{eff}^{2}\right)^{1/2},\end{equation}
Thus, the  total free energy is given by\begin{eqnarray}
F\left(\beta \right)&=&\frac{1}{2r}\sum _{n=1}^{\infty }n^{2}\left(n^{2}+\mu _{eff}^{2}\right)^{1/2}-\frac{r\mu _{eff}^{2}}{\beta ^{2}}\sum _{n=1}^{\infty }\frac{1}{n^{2}}K_{2}\left(m_{eff}\beta n\right)  \nonumber \\
&&+\frac{\mu _{eff}^{2}}{\left(2\pi \right)^{2}r}\sum _{n=1}^{\infty }\frac{1}{n^{2}}K_{2}\left(2\pi n\mu _{eff}\right)-\frac{\mu _{eff}^{3}}{2\pi r}\sum _{n=1}^{\infty }\frac{1}{n}K_{3}\left(2\pi n\mu _{eff}\right).  \label{free energy I}
\end{eqnarray}
 Exact expressions for the total energy and entropy follow from the
fundamental relations $E=d\left(\beta F\right)/d\beta $ and $S=\beta \left[d\left(\beta F\right)/d\beta -F\right]$.
The zero point contribution must, of course, be understood in terms
of its analytical continuation \cite{Elisalde94}. A shortcut that
avoids analytical continuation is the realisation that the last two
terms of Eq. (\ref{high T partition function I}) or  Eq. (\ref{free energy I}),
lead to minus the \emph{renormalised} vacuum energy,$-E_{0}$, that
is, the exact renormalised expression for the Casimir energy reads\begin{equation}
E_{0}=-\frac{\mu _{eff}^{2}}{\left(2\pi \right)^{2}}\sum _{r\, n=1}^{\infty }\frac{1}{n^{2}}K_{2}\left(2\pi n\mu _{eff}\right)+\frac{\mu _{eff}^{3}}{2\pi }\sum _{n=1}^{\infty }\frac{1}{n}K_{3}\left(2\pi n\mu _{eff}\right). 
\label{eq: exact ZTCasimir energy}\end{equation}
 For $\mu _{eff}^{2}\ll 1$, for example, we obtain \begin{equation}
E_{0}\approx \frac{1}{240r}-\frac{\mu _{eff}^{2}}{48r}-\frac{\mu _{eff}^{4}}{16r},\end{equation}
in agreement with the analytical method. Equation (\ref{eq: exact ZTCasimir energy})
is an instance of the approach to the Casimir energy proposed in
\cite{KE1996}.

\subsubsection{The high temperature approximation and small effective-mass 
corrections}

In the high-T regime, $m_{eff}\beta \ll 1$, with the effective mass
in the interval $0<m_{eff}<r^{-1}$, we can expand the modifed Bessel
functions in Eq. (\ref{high T partition function I}) to obtain \begin{equation}
\log \widetilde{Z}\left(\beta \right)\approx \frac{\pi ^{4}}{45}\frac{r^{3}}{\beta ^{3}}+\frac{\beta }{240r}-\frac{\mu _{eff}^{2}r}{2\beta }-\frac{\mu _{eff}^{2}\beta }{48r},\end{equation}
which is Kutasov and Larsen's result \cite{Kutasov and Larsen 2000}
with the lowest-order effective-mass correction. Higher-order corrections
can be easily evaluated. It follows that the thermal free energy can
be written in scaled form as \begin{equation}
r\widetilde{F}\left(\beta \right)\approx -\frac{\pi ^{4}}{45}\left(\frac{r}{\beta }\right)^{4}-\frac{1}{240}+\frac{\mu _{eff}^{2}}{2}\left(\frac{r}{\beta }\right)^{2}+\frac{\mu _{eff}^{2}}{48}.\end{equation}
 The scaled thermal energy is \begin{equation}
r\widetilde{E}\left(\beta \right)\approx \frac{\pi ^{4}}{15}\left(\frac{r}{\beta }\right)^{4}-\frac{1}{240}-\frac{\mu _{eff}^{2}}{2}\left(\frac{r}{\beta }\right)^{2}+\frac{\mu _{eff}^{2}}{48},\end{equation}
and the entropy is\begin{equation}
S\left(\beta \right)\approx \frac{4\pi ^{4}}{45}\left(\frac{r}{\beta }\right)^{3}-\mu _{eff}^{2}\frac{r}{\beta }.\end{equation}
 Again, if we set $\mu _{eff}^{2}=0$, we recover the standard results
for the conformally invariant case \cite{Klemm at al 2000,Brevik al 02}.

\subsubsection{Thermodynamical functions for $-1<\mu _{eff}^{2}<0$}

The evaluation of the logarithm of the partition function in the open
interval $-1<\mu _{eff}^{2}<0$ is somewhat more envolved. First of
all, we redefine $f\left(h\ell \right)$ according to\begin{equation}
f\left(h\ell \right)\rightarrow \widetilde{f}\left(h\ell \right)\equiv -\left(\frac{r}{\beta }\right)^{3}\left(h\ell \right)^{2}\log \left[1-e^{-\left(h^{2}\ell ^{2}-\left|m_{eff}^{2}\right|\beta ^{2}\right)^{1/2}}\right],\end{equation}
where, as before, the scale factor is $h=\beta /r$. With the replacement
$h\ell \rightarrow x$ we also have \begin{equation}
f\left(x\right)\rightarrow \widetilde{f}\left(x\right)\equiv -\left(\frac{r}{\beta }\right)^{3}x^{2}\log \left[1-e^{-\left(x^{2}-\left|m_{eff}^{2}\right|\beta ^{2}\right)^{1/2}}\right],\end{equation}
 which we will suppose to be real and hence defined in the interval
$\left|m_{eff}^{2}\right|^{1/2}\beta <x<\infty $. After introducing
the new variable $t\equiv \sqrt{x^{2}-\left|m_{eff}^{2}\right|\beta ^{2}}$,
expanding the log and treating the integral term as before the rescaled
Abel-Plana formula will read\begin{equation}
\frac{\beta }{r}\sum _{\ell =1}^{\infty }\widetilde{f}\left(h\ell \right)=-\left(\frac{r}{\beta }\right)^{3}\sum _{n=1}^{\infty }\frac{1}{n^{2}}\left[\frac{dI_{n}\left(\lambda \right)}{d\lambda }\right]_{\lambda =1}+H\left(m_{eff}\beta \right),\label{R Abel-Plana 2}\end{equation}
where we have defined\begin{equation}
I_{n}\left(\lambda \right)\equiv \int _{0}^{\infty }\, dt\, \left(t^{2}+\left|m_{eff}^{2}\right|\beta ^{2}\right)^{1/2}e^{-nt\lambda },\end{equation}
and \begin{equation}
H\left(m_{eff}\beta \right)\equiv -2\left(\frac{r}{\beta }\right)^{3}\sum _{k=1}^{\infty }\frac{1}{k}\int _{\left|m_{eff}^{2}\right|^{1/2}\beta }^{\infty }\, dy\, y^{2}\frac{\sin \left[\left(y^{2}+\left|m_{eff}^{2}\right|\beta ^{2}\right)^{1/2}\right]}{e^{2\pi ry/\beta }-1}.\end{equation}
{}From our previous experience we can anticipate that the first term
in Eq. (\ref{R Abel-Plana 2}) contains the thermal corrections and
the second one is related to the zero temperature vacuum energy. 

The integral $I_{n}\left(\lambda \right)$ can be calculated with
the help of (\emph{cf}. \cite{Grad94}, formula 3.387.7) \begin{equation}
\int _{0}^{\infty }dx\, \left(x^{2}+u^{2}\right)^{\nu -1}e^{-px}=\frac{\sqrt{\pi }}{2}\left(\frac{2u}{p}\right)^{\nu -\frac{1}{2}}\Gamma \left(\nu \right)\left[H_{\nu -\frac{1}{2}}\left(up\right)-N_{\nu -\frac{1}{2}}\left(up\right)\right],\label{Grad Int 2}\end{equation}
that holds for $\left|\arg u\right|<\pi $ and $\Re\, p>0$. The functions
$H_{\nu }\left(z\right)$ are the Struve functions whose series representation
is given by \cite{Grad94}\begin{equation}
H_{\nu }\left(z\right)=\sum _{j=0}^{\infty }\left(-1\right)^{j}\frac{\left(\frac{z}{2}\right)^{2j+\nu +1}}{\Gamma \left(j+\frac{3}{2}\right)\Gamma \left(\nu +j+\frac{3}{2}\right)},\end{equation}
 and $N_{\nu }\left(z\right)$ are the Bessel function of the second
kind (Neumann functions) \cite{Grad94,Whittaker}. It follows that
the first term of Eq. (\ref{Abel-Plana 2}) can be rewritten as\begin{equation}
\left(\frac{r}{\beta }\right)^{3}\sum _{n=1}^{\infty }\frac{1}{n^{2}}\left[\frac{dI_{n}\left(\lambda \right)}{d\lambda }\right]_{\lambda =1}=\frac{\pi r^{3}\left|m_{eff}^{2}\right|^{1/2}}{2\beta ^{2}}\sum _{n=1}^{\infty }\frac{1}{n^{3}}\left\{ \frac{d}{d\lambda }\frac{1}{\lambda }\left[H_{1}\left(\alpha \lambda \right)-N_{1}\left(\alpha \lambda \right)\right]\right\} _{\lambda =1},\end{equation}
 where $\alpha \equiv \left|m_{eff}^{2}\right|^{1/2}\beta n$. 

The second term can be integrated just as we did before, that is,
first we expand the sine function as in (\ref{sinetaylor}) and replace
the denominator using (\ref{Denominator}), then we get\begin{equation}
H\left(m_{eff}\beta \right)=\frac{r}{4\pi ^{2}\beta }\sum _{n=1}^{\infty }\frac{1}{n^{2}}\left[\frac{d^{2}}{d\lambda ^{2}}\int _{\left|m_{eff}^{2}\right|^{1/2}\beta }^{\infty }\, dy\, \left(y^{2}+\left|m_{eff}^{2}\right|\beta ^{2}\right)^{1/2}e^{-2\pi rny\lambda /\beta }\right]_{\lambda =1}.\end{equation}
This integral can be also evaluated with the help of Eq. (\ref{Grad Int 2})
and the result is\begin{equation}
\int _{\left|m_{eff}^{2}\right|^{1/2}\beta }^{\infty }\, dy\, \left(y^{2}+\left|m_{eff}^{2}\right|\beta ^{2}\right)^{1/2}e^{-2\pi rny\lambda /\beta }=\frac{\left|m_{eff}^{2}\right|^{1/2}\beta ^{2}}{4\pi rn\lambda }\times \left[H_{1}\left(\eta \lambda \right)-N_{1}\left(\eta \lambda \right)\right],\end{equation}
 where $\eta \equiv 2\pi \left|m_{eff}^{2}\right|^{1/2}rn$. Hence, we
can write\begin{eqnarray}
\log \widetilde{Z}\left(\beta \right)=-\frac{\pi r^{3}\left|m_{eff}^{2}\right|^{1/2}}{2\beta ^{2}}\sum _{n=1}^{\infty }\frac{1}{n^{3}}\left\{ \frac{d}{d\lambda }\frac{1.}{\lambda }\left[H_{1}\left(\alpha \lambda \right)-N_{1}\left(\alpha \lambda \right)\right]\right\} _{\lambda =1} &  & \nonumber \\
+\frac{\left|m_{eff}^{2}\right|^{1/2}\beta }{16\pi ^{2}}\sum _{n=1}^{\infty }\frac{1}{n^{3}}\left\{ \frac{d^{2}}{d\lambda ^{2}}\frac{1}{\lambda }\left[H_{1}\left(\eta \lambda \right)-N_{1}\left(\eta \lambda \right)\right]\right\} _{\lambda =1}. &  & \label{log partial}
\end{eqnarray}
Equation (\ref{log partial}) can be simplified by making use of a
relation similar to Eq. (\ref{recursion 1}) for the Neumann functions
and of the relations \cite{Grad94}\begin{equation}
\frac{d}{dz}z^{-\nu }H_{\nu }\left(z\right)=\frac{1}{2^{\nu }\pi ^{1/2}\Gamma \left(\nu +\frac{3}{2}\right)}-z^{-\nu }H_{\nu +1}\left(z\right),\end{equation}
and\begin{equation}
2\frac{d}{dz}H_{\nu }\left(z\right)=H_{\nu -1}\left(z\right)-H_{\nu +1}\left(z\right)+\frac{\left(\frac{1}{2}z\right)^{\nu }}{\pi ^{1/2}\Gamma \left(\nu +\frac{3}{2}\right)},\end{equation}
Using these relations, we can recast Eq. (\ref{log partial}) into
the form\begin{eqnarray}
\log \widetilde{Z}\left(\beta \right)=-\frac{\pi r^{3}\left|m_{eff}^{2}\right|^{1/2}}{2\beta ^{2}}\sum _{n=1}^{\infty }\frac{1}{n^{3}}\left[\frac{2\alpha ^{2}}{3\pi }-\alpha H_{2}\left(\alpha \right)+\alpha N_{2}\left(\alpha \right)\right] &  & \nonumber \\
+\frac{\left|m_{eff}^{2}\right|^{1/2}\beta }{16\pi ^{2}}\sum _{n=1}^{\infty }\frac{1}{n^{3}}\left\{ -\eta ^{3}\left[\frac{H_{1}\left(\eta \right)}{2\eta }-\frac{H_{2}\left(\eta \right)}{\eta ^{2}}-\frac{H_{3}\left(\eta \right)}{2\eta }+\frac{\eta }{60\pi }\right]+\eta N_{2}\left(\eta \right)-\eta ^{2}N_{3}\left(\eta \right)\right\} _{\lambda =1}. &  & \label{log partial 2}
\end{eqnarray}
The first term in the first line on the r.h.s. of Eq. (\ref{log partial 2})
leads to the divergent parcel $-\left|\mu _{eff}^{2}\right|^{3/2}\zeta \left(1\right)/3$,
nevertheless, this contribution does not depend on $\beta $ or $r$,
thus being physically unobservable. Therefore, we can discard it without
too much ado, and with the deletion of this term, Eq. (\ref{log partial 2})
is our final result. As a check consider again the massless limit.
Since the parameters $\alpha $ and $\eta $ are proportional to $\left|m_{eff}^{2}\right|^{1/2}$,
and the Struve functions are represented by ascending powers of $\alpha $
and $\eta $, their contribution in the massless limit is negligible.
The same remark holds for the quartic term in $\eta $. Then, recalling
that for small $z$ the limiting form of Neumann functions is $N_{\nu }\left(z\right)\simeq -\left(1/\pi \right)\Gamma \left(\nu \right)\left(z/2\right)^{-\nu }$,
we actually obtain Kutasov and Larsen's result \cite{Kutasov and Larsen 2000}. 

The regularised vacuum energy for the open interval $-1<\mu _{eff}^{2}<0$
can be inferred from Eq. (\ref{log partial 2}) and reads\begin{equation}
E_{0}=\frac{\left|\mu _{eff}^{2}\right|^{1/2}}{16\pi ^{2}r}\sum _{n=1}^{\infty }\frac{1}{n^{3}}\left\{ \eta ^{3}\left[\frac{H_{1}\left(\eta \right)}{2\eta }-\frac{H_{2}\left(\eta \right)}{\eta ^{2}}-\frac{H_{3}\left(\eta \right)}{2\eta }+\frac{\eta }{60\pi }\right]-\eta N_{2}\left(\eta \right)+\eta ^{2}N_{3}\left(\eta \right)\right\} _{\lambda =1}.\end{equation}
Effective mass corrections are, as before, easily obtainable. For
example, to lowest order in $\left|\mu _{eff}^{2}\right|$, the logarithm
of the partition function is given by \begin{equation}
\log \widetilde{Z}\left(\beta \right)\approx \frac{\pi ^{4}}{45}\frac{r^{3}}{\beta ^{3}}+\frac{\beta }{240r}+\frac{\pi ^{2}\left|\mu _{eff}^{2}\right|}{12}\frac{r}{\beta }+\frac{3\left|\mu _{eff}^{2}\right|}{16}\frac{\beta }{r}.\end{equation}
 The relevant thermodynamical functions in this approximation read\begin{equation}
r\widetilde{E}\left(\beta \right)\approx \frac{\pi ^{4}}{15}\frac{r^{4}}{\beta ^{4}}-\frac{1}{240}+\frac{\pi ^{2}\left|\mu _{eff}^{2}\right|}{12}\frac{r^{2}}{\beta ^{2}}-\frac{3\left|\mu _{eff}^{2}\right|}{16},\end{equation}
\begin{equation}
r\widetilde{F}\left(\beta \right)\approx -\frac{\pi ^{4}}{45}\frac{r^{4}}{\beta ^{4}}-\frac{1}{240}-\frac{\pi ^{2}\left|\mu _{eff}^{2}\right|}{12}\frac{r^{2}}{\beta ^{2}}-\frac{3\left|\mu _{eff}^{2}\right|}{16},\end{equation}
\begin{equation}
S\left(\beta \right)\approx \frac{4\pi ^{4}}{45}\frac{r^{3}}{\beta ^{3}}+\frac{\pi ^{2}\left|\mu _{eff}^{2}\right|}{6}\frac{r}{\beta }.\end{equation}
Notice that the entropy is always positive, but that, owing to quantum effects,
the energy can attain negative values. Higher order corrections include
fractional powers of $\left|\mu _{eff}^{2}\right|$. It is important,
however, to realise once again that, in the thermodynamical sense, the
conformal limit has in our approach a broader sense.

\section{Entropy/energy ratios}

Now we can evaluate the entropy/energy ratios. Though they are meaningful
for the high-temperature regime only, for the sake of completness we will
 also evaluate
them in the low-temperature limit. To start, in the low temperature regime the
entropy/energy ratio is $\mu _{eff}$-independent. In fact, from Eqs.
(\ref{lowTentropy}) and (\ref{lowTenergy}) we readily obtain \begin{equation}
\frac{S\left(\beta \right)}{2\pi r\widetilde{E}\left(\beta \right)}=\delta \gg 1,\end{equation}
where, for comparison with known results, we have introduced the notation
$\delta \equiv \beta /\left(2\pi r\right)$. This is the massless conformally
invariant result \cite{Brevik al 02}. 

The entropy/energy ratio in the high-temperature regime and for 
$0<\mu _{eff}^{2}<1$
is given by\begin{equation}
\frac{S\left(\beta \right)}{2\pi r\widetilde{E}\left(\beta \right)}\approx \frac{\frac{\delta ^{-3}}{180}-\frac{\mu _{eff}^{2}}{8\pi ^{2}}\delta ^{-1}}{\left[\frac{1}{240}\left(\delta ^{-4}-1\right)-\frac{\mu _{eff}^{2}}{8\pi ^{2}}\left(\delta ^{-2}-\frac{\pi ^{2}}{6}\right)\right]},\, \, \, \, \delta \ll 1.\label{ratio one}\end{equation}
Again, by setting $\mu _{eff}^{2}=0$ we recover the conformal case
result \cite{Brevik al 02,Klemm at al 2000}, but this time extended
to the non-conformal field. 

The entropy/energy ratio in the high-temperature regime with
 $-1<\mu _{eff}^{2}<0$,
reads\begin{equation}
\frac{S\left(\beta \right)}{2\pi r\widetilde{E}\left(\beta \right)}\approx \frac{\frac{\delta ^{-3}}{180}+\frac{\left|\mu _{eff}^{2}\right|}{24}\delta ^{-1}}{\left[\frac{1}{240}\left(\delta ^{-4}-1\right)+\frac{\left|\mu _{eff}^{2}\right|}{48}\left(\delta ^{-2}-9\right)\right]},\, \, \, \, \delta \ll 1.\label{ratio two}\end{equation}
These ratios can be compared and visualised in the plot below (Fig. 1), where
we also show the Bekenstein and the Verlinde ratios
\begin{figure}[thb]
\vspace{0.3cm}
\begin{center}\includegraphics[  width=8cm,
  height=8cm]{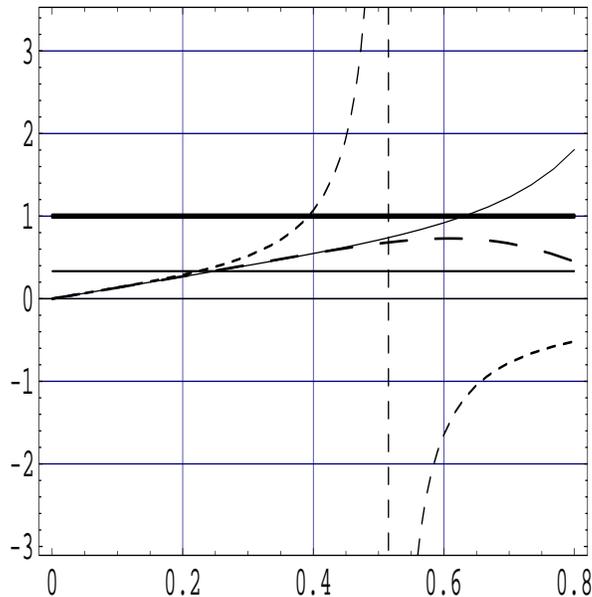}\end{center}

\caption{\protect\small The entropy/energy ratios for our configuration,
in the two different regimes of $\mu _{eff}$, namely $0<\mu _{eff}^{2}<1$
 (long dash curve) and $-1<\mu _{eff}^{2}<0$  (short dash curve), as 
compared with the conformal case (thin full line) and with the  Bekenstein 
and Verlinde  bounds (thicker line and second thicker one,
respectively).}
\end{figure}

The horizontal axis is $\delta $ and the vertical one is the ratio
entropy/energy. The thicker line  is the upper limit of Bekenstein's ratio 
$S\left(\beta \right)/2\pi r\widetilde{E}\left(\beta \right)=1$.
 The thinner of the thick lines is the Verlinde ratio
 $S\left(\beta \right)/2\pi r\widetilde{E}\left(\beta 
\right)=1/3$. The thin full curve is the conformal result. The
long-dashed curve plot is Eq. (\ref{ratio one}) and the short-dashed
one Eq. (\ref{ratio two}); in both cases we have set 
$\mu _{eff}^{2}=\left|\mu _{eff}^{2}\right|=0.5$.
For $\delta \rightarrow 0$ the plots corresponding to Eqs. (\ref{ratio one}),
(\ref{ratio two}), and the conformal case all coalesce into the same
straight line. Notice that for $\delta \approx 0.2$ all entropy bounds
exceed the Verlinde bound. The one corresponding to the conformal limit 
surpases quite soon the Bekenstein bound, at about $\delta \approx 0.6$.
Concerning our approach, it is very interesting to remark that,
owing to the different algebraic signs of the effective mass corrections,
the entropy bounds given by Eqs. (\ref{ratio one}) and (\ref{ratio two})
behave in a very different way. Indeed, with negative (tachyonic) effective
mass, the entropy bound soon exceeds ($\delta \approx 0.4$) the
Bekenstein bound (as in the case of the conformal theory). Here, this 
is probably the manifestation of some tachyonic-related
instability, that renders also another branch of negative values of the 
ratio. Notice the sudden jump for $\delta \approx 0.5$. This
behaviour also seems to be related to such tachyonic instabilituy
and is due to the fact that the energy  becomes 
negative. The entropy
remains always positive, as remarked before. However, because we are
on the verge of violating the validity of our approximation the behaviour
for $\delta > 0.5$ is suspicious and could be spurious. At the same
time, for {\it positive effective mass} the entropy bound remains below
the Bekenstein bound for {\it all} values of $\delta$, which is a remarkable
result. What is more, the value seems to come back to fulfill Verlinde's 
bound, for higher values of $\delta$, which is indeed very nice.

{}From Eq. (\ref{high T partition function I}) we can also obtain the
entropy/energy ratio in the limit $m_{eff}\beta \gg 1$. In fact,
it is easy to show that in this limit the entropy/energy ratio does
not depend on the effective mass $m_{eff}$ and reads\begin{equation}
\frac{S\left(\beta \right)}{2\pi r\widetilde{E}\left(\beta \right)}\approx 
\frac{5}{3}\delta \, .\end{equation}
Therefore, in the high-temperature regime ($\delta \rightarrow 0)$,
with $m_{eff}\gg T$, the ratio entropy/energy remains bounded by
the Bekenstein limit. It should be expected that a similar bound will hold
for the de Sitter graviton in the early thermal universe.

\section{Concluding remarks}

In this work, we have considered the evaluation of of the Casimir energy
at zero and finite temperature for the case of a massive scalar field
in $\mathbf{S}^{1}\times \mathbf{S}^{3}$ geometries allowing 
for an arbitrary conformal parameter $\xi $.

The evaluation of the corresponding thermodynamical functions through
the rescaled Abel-Plana sum formula leads straightforwardly to the
very high temperature regime (no exponential corrections), but then,
it is there that the relevant physics lies. As stressed in Ref. \cite{Brevik al 02},
it is only in the high-T regime that we are allowed to make use of
thermodynamical methods in the one-loop approximation \cite{Dasetal2002}
(for earlier studies of similar questions see Ref. \cite{Dowker88}
where a slightly different approach was employed). We have also shown
that for $\mu _{eff}^{2}=0$, the zero temperature Casimir energy
is degenerated in the sense that it is equal to the Casimir energy
of the conformal case. We also have evaluated the relevant thermodynamical
functions and extended the previous analysis of entropy bounds and
entropy-energy ratios of the symmetrical conformal case to the present
one. Interestingly enough, for $\mu _{eff}^{2}=0$, all previous results
concerning Bekenstein-Verlinde ratios are valid. Also, dual symmetries,
as for example the temperature inversion symmetry, are recovered. 

Our results can be used in the calculation of the entropy/energy
ratio for the graviton living in the same spacetime. Since the cosmological
constant will play the role of the effective mass considered here, it 
follows that
the entropy/energy ratio will depend on the cosmological constant.
Moreover, since a `cosmological constant' term coming from the 
vacuum fluctuations of some field (as predicted by several 
fashionable models) will depend on time,  as the radius of the
universe also does, this bound will become time-dependent. We
can speculate then on the relation between the change of such a kind of
 cosmological constant and the evolution of the entropy bound. 

As a final remark, notice that the free energy calculated here can
be used in the analysis of the back reaction of a massive scalar field
in the spacetime under discussion.

\subsection*{Acknowledgments}

The authors are indebted with Sergei Odintsov for important suggestions 
and very helpful comments.
A. C. T. wishes to acknowledge the kind hospitality of the Institut
d'Estudis Espacials de Catalunya (IEEC/CSIC) and the Universitat de
Barcelona, Departament d'Estructura i Constituents de la Mat\`{e}ria.
A. C. T. also acknowledges the financial support of Consejo Superior
de Investigaci\'on Cient\'{\i}fica (CSIC, Spain), and of CAPES,
the Brazilian agency for faculty improvement, Grant BExt 06832/01-2.
The investigation of E.E. has been supported by DGI/SGPI (Spain),
project BFM2000-0810, and by CIRIT (Generalitat de Catalunya), contracts
2002BEAI400019, 2001ACES00014 and 2001SGR-00427.

\end{document}